\documentclass[runningheads]{llncs}

\usepackage[a-1b]{pdfx}
\usepackage{hyperref}

\usepackage{booktabs} 
\usepackage{caption}
\usepackage{mathtools} 
\usepackage{setspace} 
\usepackage{float}
\usepackage{graphicx}
\usepackage[colorinlistoftodos]{todonotes}
\usepackage{url}
\makeatletter
\g@addto@macro{\UrlBreaks}{\UrlOrds}
\makeatother
\DeclarePairedDelimiter{\norm}{\lVert}{\rVert}
\usepackage[normalem]{ulem}
\useunder{\uline}{\ul}{}

\begin{document}
\title{Leveraging Schema Labels \\to Enhance Dataset Search}
\author{Zhiyu Chen \and
Haiyan Jia   \and \\
Jeff Heflin \and
Brian D.\ Davison }
\authorrunning{Chen et al.}


\institute{Lehigh University, Bethlehem, PA, USA\\
\email{\{zhc415,haiyan.jia\}@lehigh.edu}\\
\email{\{heflin,davison\}@cse.lehigh.edu}\\
}

\maketitle
\pagestyle{empty}

\begin{abstract}

A search engine's ability to retrieve desirable datasets is important for data sharing and reuse. Existing dataset search engines typically rely on matching queries to dataset descriptions. However, a user may not have enough prior knowledge to write a query using terms that match with description text.
We propose a novel schema label generation model which generates possible schema labels based on dataset table content. We incorporate the generated schema labels into a mixed ranking model which not only considers the relevance between the query and dataset metadata but also the similarity between the query and generated schema labels. To evaluate our method on real-world datasets, we create a new benchmark specifically for the dataset retrieval task. Experiments show that our approach can effectively improve the precision and NDCG scores of the dataset retrieval task compared with baseline methods. We also test on a collection of Wikipedia tables to show that the features generated from schema labels can improve the unsupervised and supervised web table retrieval task as well.

\end{abstract}

\section{Introduction}

Dataset retrieval is receiving more attention as people from different fields and domains start to rely on datasets for their work.
There are many data portals with the purpose of effective and efficient data management and data sharing, such as data.gov\footnote{\label{data.gov}\url{https://www.data.gov/}}, 
 datahub\footnote{\url{http://datahub.io/}} and data.world\footnote{\url{https://data.world/}}.  
 Most of those data portals use  CKAN\footnote{\url{https://docs.ckan.org/}} as their backend. 
 However, there are two problems of dataset search engines using such infrastructure: First, ranking performance relies on the quality of metadata of datasets, while many datasets lack high quality metadata; second, the information in the metadata may not satisfy the user's information need or help them solve their task \cite{chapman2019dataset}. A user may not know the organization of a potentially relevant dataset, or the tags data publishers provide with a dataset. Such information can hardly be used for dataset ranking. 

\begin{figure}[t]
\centering
\includegraphics[width=0.95\textwidth]{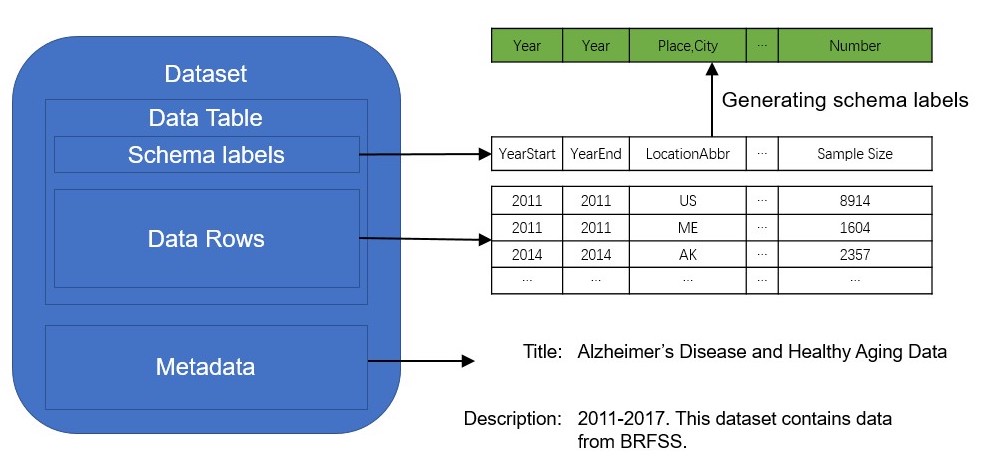}
\small
\caption{The structure of a dataset.  Metadata includes the title and any description. A trained schema label generator is used to generate additional schema labels (green part) from similar data tables.}
\label{dataset}
\end{figure}

In this paper, we focus on the problem of dataset retrieval where dataset content is in tabular form, since tabular data  
is widely-used and easy to read and write. 
As illustrated in Fig.~\ref{dataset}, a dataset consists of a data table (dataset content) and metadata. A data table usually has one header row, followed by one or more data rows. The header row consists of a list of \textbf{schema labels} (attribute names) whose actual values are stored in data rows. Metadata usually includes title and description of the dataset.

Schema labels, which represent high-level concepts, are underutilized if we directly score them with a user query. 
Consider the example in Fig.~\ref{dataset}; the  vocabulary of schema labels could be very different from other fields and user queries. ``LocationAbbr'', standing for ``Location Abbreviation'', is unlikely to appear in a user query so this dataset is less likely to be recalled. 
However, we can enhance this dataset by generating schema labels such as ``place'' and ``city'' appearing in other, similar datasets, which could provide a better soft-matching signal with respect to a user query, and therefore increase the chance that it can be recalled.

In this work, we first propose a new method for schema label generation. We learn latent feature representations of schema labels automatically by jointly decomposing the dataset-schema label interaction matrix and schema label-schema label interaction matrix.
Then we propose a framework for enhancing dataset retrieval by schema label generation to address the problem that schema labels are not effectively used by existing dataset search engines.
We create a new public benchmark\footnote{Available via  \url{https://github.com/Zhiyu-Chen/ECIR2020-dataset-search} } based on federal (U.S.) datasets 
and use it to demonstrate the effectiveness of our proposed framework for dataset retrieval. 
We additionally consider a web table retrieval task and demonstrate that the features generated from schema labels can be effective for supervised ranking.

\section{Related Work}

Dataset search has become a new research field with new challenges.  Chapman et al.~\cite{chapman2019dataset}  classify dataset search into {\em basic} and {\em constructive} dataset search. Basic dataset search returns a list of existing datasets based on a user's query, while constructive dataset search \cite{gentile2016extending} generates datasets on-the-fly based on a user's needs and query.
Google recently released a dataset search service\footnote{\url{https://toolbox.google.com/datasetsearch}}. Like many other data portals, their service relies on metadata of datasets, annotated on web pages using a standard defined by schema.org.

Other work on applications of Web tables is also related to our work.
Cafarella et al.~\cite{cafarella2008webtables} proposed WebTables system which extract Web tables from top ranked pages by keyword search.
Sekhavat et al.~\cite{sekhavat2014knowledge} proposed a probabilistic method that augments an existing knowledge base with facts from Web tables. 
Zhang et al.~\cite{Zhang:2017:ESA} developed generative probabilistic models to equip spreadsheets with smart assistance capabilities. Specifically, given a table, they recommend additional rows and column headings by leveraging the information from the Web tables.   They also developed semantic matching features for table retrieval~\cite{zhang2018ad}. 

The techniques designed for Web table analysis could potentially be applied to dataset search. In our work, each dataset is associated with data in tabular form. Extracting useful information from tables such as entities and attribute names could help with the retrieval task. Trabelsi et al.~\cite{mo2019bdt}  recently proposed custom embeddings for column headers based on multiple contexts for table retrieval, and found representing numerical cell values to be useful.  
Zhang et al.~\cite{Zhang:2017:ESA} proposed to use semantic concepts to represent queries and tables for ranking entity-focused tables.  However, dataset search could be inherently more difficult since datasets do not need to be entity-focused.

\section{Schema Label Enhanced Ranking}

\begin{figure*}[h]
\centering
\includegraphics[width=0.95\textwidth]{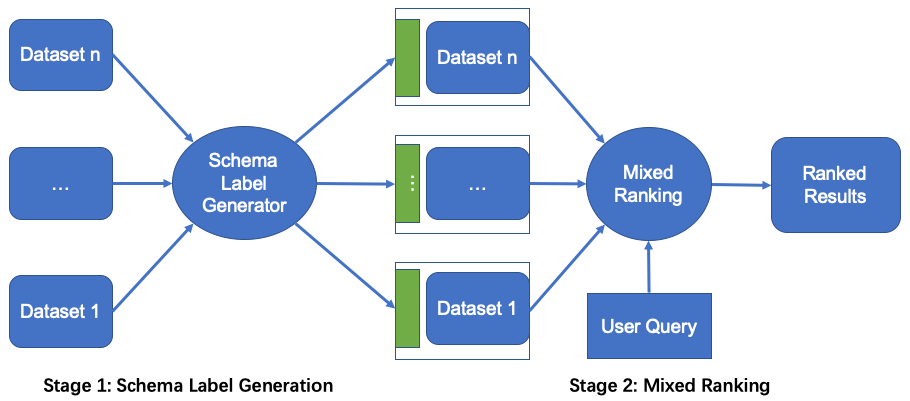}
\caption{The proposed schema label enhanced dataset retrieval framework. The green blocks indicate generated schema labels for different datasets.}
\label{framework}
\end{figure*}

In this section, we introduce the framework of schema label enhanced dataset retrieval. As illustrated in Fig.~\ref{framework}, our framework has two stages: in the first stage, we first train a schema label generator with the method proposed in Section~\ref{slg} and use it to generate additional schema labels for all the datasets;
in the second stage, we use a mixed ranking model to combine the scores of schema labels and other fields for dataset ranking.
In the following subsections, we present a detailed illustration of the two stages.

\subsection{Schema Label Generation}\label{slg}

We propose to improve dataset search by making use of generated schema labels, since these can be complementary to the original schema labels and especially valuable when they are otherwise absent from a dataset.

We treat schema label generation as a multi-label classification problem.  Let  $L = \{l_1,l_2,...,l_k\}$  denote the labels appearing in all datasets and $D = \{  (\mathbf{x}^i,\mathbf{y}^i) | 1 \leq i \leq n\}$  denote the training set. Here, for each training sample $(\mathbf{x}^i,\mathbf{y}^i)$, $\mathbf{x}^i$ is a d-dimensional feature vector of column $i$ which can be calculated from data rows~\cite{chen2018generating} or learned from matrix factorization proposed later in this section.  $\mathbf{y}^i$ is k-dimensional vector $[y_1^i, y_2^i,...,y_k^i]$  and $y_j^i = 1$ only if $x_i$ is relevant to label $l_j$, otherwise  $y_j^i = 0$.
Our objective is to learn a function that models $P(l|x_i)$, $(l\in L)$. To generate $m$ schema labels for column $i$, we can select the top $m$ labels $L_m $ by: 
$$L_m = \operatorname*{arg\,max}_{ l\in L_m \subseteq  L} P(l|x_i)$$
We could also generate schema labels by selecting a probability threshold $\theta$:
$$L_m = \{l \in L | P(l|x_i) \geq \theta \}$$
In practice, we could first generate the top $m$ schema labels and filter out those results with a probability lower than the threshold.

Chen et al.~\cite{chen2018generating} proposed to predict schema labels based on curated features of data values. 
Instead of designing curated features for schema labels, we consider learning their representations in an automated manner. Inspired by collaborative filtering methods in recommender systems, we model each dataset as a user and each schema label as an item. Then a dataset with a schema label can be considered as positive feedback between a user and an item.  By exploiting the user-item co-occurrences and item-item co-occurrences, we can learn the latent representations of schema labels. In the following, we show how to construct a preference matrix in the context of schema label generation and how to learn the schema label features.

\textbf{Preference Matrix Construction.} 
With $m$ data tables and $n$ unique schema labels, we can construct a dataset-column preference matrix $M^{m \times n}$, where $M_{up}$ is 1 if dataset $u$ contains schema label $p$. 

\textbf{Matrix Factorization.} 
MF \cite{koren2009matrix} decomposes $M$ into the product of $U^{m \times k}$ and $P^{k \times n}$ where $k < min(m,n)$. $U^T$ can be denoted as $ (\alpha_1, ..., \alpha_u\, ..., \alpha_m)$ where $\alpha_u \in R^k $ represents the latent factor vector of dataset $u$. Similarly, $P^T$ can be denoted as $ (\beta_1, ..., \beta_p\, ..., \beta_n)$ where $\beta_p \in R^k$ represents the latent factor vector of schema label $p$.
Since the preference matrix actually models the implicit feedback, MF optimizes the following objective function:
\begin{equation}
\mathcal{L}_{mf} = \sum_{u,p}c_{up}(M_{up} - \alpha_u^{T} \beta_{p})^2 + \lambda_\alpha \sum_u \norm{\alpha_u}^2  + \lambda_\beta \sum_p \norm{\beta_p}^2
\end{equation}
where $c_{up}$ is a hyperparameter tuned to balance the non-zero and zero values  since $M$ is a sparse matrix. $\lambda_\alpha $ and $\lambda_\beta$ are regularization parameters that adjust the importance of regularization terms $ \sum_u \norm{\alpha_u}^2$ and $\sum_p \norm{\beta_p}^2$.

\textbf{Label Embedding.}  
Recently, word embedding techniques (e.g., word2vec \cite{mikolov2013distributed}) have been valuable in natural language processing tasks. Given a sequence of words, a low-dimensional continuous representation called word embedding can be learned for each word.  Word2vec's skip-gram model with negative sampling (SGNS) is equivalent to implicitly factorizing a word-context matrix, whose cells are the pointwise mutual information (PMI) of the respective word and context pairs, shifted by a global constant \cite{levy2014neural}.  
The PMI between word $i$ and its context word $j$ is defined as:
$$PMI(i,j) = log\frac{P(i,j)}{P(i) \times P(j)} = log \frac{\#(i,j) \times|D|}{\sum_j \#(i,j) \times \sum_i\#(i,j)}$$
where $\#(i,j)$ is the number of times word $j$ appears in the context window of word $i$ and $|D|$ is the total number of word-context pairs.
Then, a shifted positive PMI (SPPMI) of word $i$ and word $j$ is calculated as: 
\begin{equation} \label{sspmi}
SSPMI(i,j) = max\{PMI(i,j) - log(k),0\}
\end{equation}
where $k$ is the number of negative samples of SGNS.  Given a corpus,  matrix $M^{SPPMI}$ can be constructed  based on equation  \eqref{sspmi} and factorizing it is equivalent to performing SGNS.

A schema label exists in the context of other schema labels.
Therefore, we 
perform word embedding techniques to learn the latent representations of schema labels. However, we do not consider the order of schema labels. Therefore, given a schema label, all other schema labels which come from the same data table are considered as its context. With the constructed SSPMI matrix of co-occurring schema labels, we are able to decompose it to learn the latent representations of schema labels. 

\textbf{Joint Learning of Schema Label Representations.} 
Schema label representations learned from MF capture the interactive information between datasets and schema labels, while the word2vec style representations explain the co-occurrence relationships of schema labels. We 
use the CoFactor model \cite{liang2016factorization} to jointly learn schema label representations from both dataset-label interaction and label-label interaction:
\begin{equation}
\label{loss}
\begin{aligned}
\mathcal{L} &=\overbrace{ \sum_{u,p}c_{up}(M_{up} - \alpha_u^{T} \beta_{p})^2 }^{MF} \\
&+ \overbrace{\sum_{M_{pi}^{SPPMI} \neq 0}(M_{pi}^{SPPMI} - \beta_p^T \gamma_i - b_p - c_i)^2}^\text{\em schema\ label\ embedding} \\
&+ \lambda_\alpha \sum_u \norm{\alpha_u}^2  + \lambda_\beta \sum_p \norm{\beta_p}^2 +\lambda_\gamma \sum_i \norm{\gamma_i}^2 
\end{aligned}
\end{equation}
From the objective function we can see the schema label representation $\beta_p$ is shared  between MF and schema label embedding. $\gamma_i$ is the latent representation of context embedding. $b_p$ and $c_i$ are the schema label embedding bias and context embedding bias, respectively.  The last line of Equation \ref{loss} incorporates regularization terms with different $\lambda$ controlling their effects.
We use the vector-wise ALS algorithm \cite{yu2014parallel} to optimize the parameters.

\textbf{Schema label generation.}
After obtaining the jointly learned representations of schema labels, we can use them as features for schema label generation.
In this paper, we use the concatenation of schema label representations introduced here and the curated features proposed by Chen et al.~\cite{chen2018generating} to construct each $x^i$.  Any multi-label classification models can be used to train the schema label generator and in this paper we choose Random Forest.

\subsection{The Mixture Ranking Model} 

Based on the schema label generation method proposed above, we index the generated schema labels for each dataset. Now, each dataset has the following fields: metadata, data rows, schema labels and generated schema labels. 
A straightforward way to rank datasets is to use traditional ranking methods for documents.  

Zhang and Balog \cite{zhang2018ad} represent tables as single field documents or multifield documents for table retrieval task.  
For \textit{single field document representation}, a dataset is treated as a single document by concatenating the text from all the fields. Then traditional methods such as BM25 can be used to to score the dataset. For \textit{multifield document representation}, each field is scored independently against the query and a weighted sum is used for ranking. 



In our \textbf{Schema Label Mixed Ranking (SLMR)} model, we score schema labels differently from other fields. The focus of our work is to learn how schema labels, data rows and other metadata may differently influence dataset retrieval performance. Note that, for simplicity, we consider the other metadata (title and description) as a single text field, since title and description are homogeneous compared with schema labels and data rows. Therefore, we have the following scoring function for a dataset $D$:
\begin{equation}
\label{mix}
score(q,D) =  \sum_{ i \in \{text,data\}}w_{i} \times score_{text}(q,F_{i})  + w_{l} \times score_{l}(q,F_{l})  
\end{equation}
where $F_{text}$ denotes the concatenation of title and description, $F_{data}$ denotes the data table, and $F_{l}$ denotes the generated schema labels. Each field has a corresponding weights. $F_{text}$ and $F_{data}$  have the same scoring function $score_{text}$ while $F_{l}$ has a different scoring function $score_{l}$.  For $F_{text}$ and $F_{data}$, we can use a standard scoring function for normal documents. In the experiments, we use BM25 as $score_{text}$.


Due to the existence of a large number of non-dictionary words in schema labels \cite{chen2018generating} that would otherwise be outside of the vocabulary of a word-based embedding, we represent schema labels and query terms using fastText \cite{bojanowski2017enriching} in $score_{l}$, since such word embeddings are calculated from character n-grams instead of terms.
To score the schema labels with respect to a query, we use the negative Word Mover's Distance (WMD) \cite{kusner2015word}. 
WMD measures the dissimilarity between two text documents as the minimum amount of distance that the word embeddings of one document need to ``travel'' to reach the word embeddings of another document. So $score_{l}(q,F_{l}) = - wmd(fasttext(q),fasttext(F_{l}))$ reflects the semantic similarity between a query and schema labels.

\section{Data Collection}
Here we describe how we construct the new benchmark for dataset retrieval in detail. We collected 2417 resources published by the U.S.\ federal government from Data.gov which cover a variety of topics. Each resource includes one or more CSV format data tables and corresponding metadata. Each CSV table is treated as a single dataset and we use the resource-level metadata to annotate each dataset.

\subsection{Task Creation and Query Collection}

We created six tasks in which each describes a separate information need to find one or more datasets. For each, we have a statement about the information need which describes what datasets are considered as relevant. We additionally verified for each task  
the existence of at least one relevant dataset.
The dataset is public available\footnote{Available from \url{https://github.com/Zhiyu-Chen/ECIR2020-dataset-search} }.  

We used Amazon Mechanical Turk\footnote{\url{https://www.mturk.com/}} to obtain diverse queries for these tasks from real users.
Every annotator was presented with the task descriptions and asked to provide a query for each created task. To avoid the impact of task order on the quality of annotations, we randomly shuffled the order of tasks for each annotator.
We paid one dollar for each completed annotation job and 20 queries were collected for each task.
Every collected query was manually examined and obviously unrelated queries were excluded from the collection.

\subsection{Relevance Assessments}

For each task and each suggested query, we used traditional ranking functions to score single field representations of each dataset and collect the top 100 results. The following ranking models were used: BM25, TF-IDF, Language model based on Jelinek-Mercer smoothing, and Language Model with Dirichlet smoothing. 
We also used each model with two different representations:
the concatenation of all fields of the dataset and the concatenation of title and description. 
This leads to eight baselines for the pooled results.

Then, the collected task-dataset pairs were annotated for relevance using the crowdsourcing service provided by Figure Eight\footnote{\url{https://www.figure-eight.com/}}.
We did not annotate the {\em query}-dataset pairs because the goal of dataset retrieval is to find relevant datasets with respect to a {\em task} which represents the real information need.

Annotators were presented with the task title, description and link to the data table. 
Each task-dataset pair was judged on a four point scale: 0 (off topic), 1 (poor), 2 (good), and 3 (excellent).\footnote{The following labeling guidance was provided to annotators: {\em a dataset is off topic if the information does not satisfy the information need, and should not be listed in the search results from a search engine; a dataset is poor if a search engine were to include this in the search results, but it should not be listed at the top; a dataset is good if you would expect this dataset to be included in the search results from a search engine; a dataset is excellent if you would expect this dataset ranked near the top of the search results from a search engine.}}  Every annotator was paid 10 cents per task-dataset judgement.

Every single task-dataset pair was judged by three annotators and we take the majority vote as the relevance label. If no majority agreement is achieved, we take the average of the scores as the final label. The statistics of annotation results is shown in Table \ref{annotate}.

\begin{table}[tb]\footnotesize
\centering
\caption{For each task, the number of pairs assigned to each relevance label. }
\small
\label{annotate}
\begin{tabular}{@{}ccccc@{}}
\toprule
Task \# & off topic & poor & good & excellent \\ \midrule
1 & 1006 & 34 & 37 & 64 \\
2 & 164 & 248 & 585 & 308 \\
3 & 300 & 270 & 456 & 153 \\
4 & 246 & 324 & 660 & 289 \\
5 & 162 & 246 & 355 & 198 \\
6 & 181 & 303 & 614 & 367 \\ \bottomrule
\end{tabular}
\end{table}

\begin{table}[!htbp]
\centering
\small
\caption{NDCG@k and Precision@k of different models on dataset retrieval. 
The superscript + shows statistically significant improvements for our 
SLMR model over other single and multifield document ranking models. T means title, D means description, DT means data table, G means generated schema labels. }
\scriptsize
\label{ndcg}
\begin{tabular}{@{}llllllllll@{}}
\toprule
\multicolumn{1}{c}{Method} & \multicolumn{1}{c}{Used Fields} & \multicolumn{1}{c}{NDCG@5} & \multicolumn{1}{c}{@10} & \multicolumn{1}{c}{@20} & \multicolumn{1}{c}{@50} & \multicolumn{1}{c}{P@5} & \multicolumn{1}{c}{@10} & \multicolumn{1}{c}{@20} & \multicolumn{1}{c}{@50} \\ \midrule
SDR & T+D & 0.8920 & 0.8490 & 0.8222 & 0.8121 & 0.4122 & 0.3652 & 0.3452 & 0.3585 \\
SDR & DT & 0.7378 & 0.7036 & 0.6964 & 0.7107 & 0.2856 & 0.2974 & 0.2931 & 0.3122 \\
SDR & T+D+DT & 0.8435 & 0.7954 & 0.7763 & 0.7785 & 0.2574 & 0.2870 & 0.3170 & 0.3357 \\
MDR & T+D+DT & 0.9285 & 0.8874 & 0.8683 & 0.8631 & 0.4086 & 0.3612 & \bf 0.4026 & 0.3767 \\
SLMR & T+D+G & \bf 0.9293\textsuperscript{+} & \bf 0.8898 & \bf 0.8722\textsuperscript{+} & \bf 0.8662 & \bf 0.5000\textsuperscript{+} & \bf 0.4388\textsuperscript{+} & 0.4000 & 0.3761 \\
SLMR & T+D+DT+G & 0.9169 & 0.8808 & 0.8680 & 0.8555 & \bf 0.5000\textsuperscript{+} & 0.4345\textsuperscript{+} & 0.4013 & \bf 0.3783 \\ \bottomrule
\end{tabular}
\end{table}

\section{Evaluation}

\subsection{Evaluation Metrics }

We evaluate dataset retrieval performance over a range of metrics: Precision at $k$ and Normalized Discounted Cumulative Gain (NDCG) at $k$ \cite{jarvelin2002cumulated}.
To test the significance of differences between model performances, we use paired t-tests with significance at the $p=0.01$ level.

\subsection{Baselines}

We first present the baseline retrieval methods.

\textbf{Single-field document ranking (SDR).} A dataset is considered as a single document. We use BM25 to score the concatenation of title and description, the text of the data table and the concatenation of all of them. By comparing the three results, we can learn about field level importance for dataset retrieval. Parameters are chosen by grid search.

\textbf{Multifield document ranking (MDR).} By setting $w_l =0$, Eq.~\eqref{mix} degenerates to the Mixture of Language Models~\cite{ogilvie2003combining}. BM25 is also used here as $score_{text}()$ in order to have a fair comparison with other methods. To optimize field weights, we use coordinate ascent. Finally, smoothing parameters are optimized in the same manner as single-field document ranking.

\subsection{Experimental Results}

In this section, we examine the following research questions:

\begin{itemize}
    \item[\textbf{Q1}] Does data table content help in dataset retrieval?
    \item[\textbf{Q2}] Do generated schema labels help in dataset retrieval?
    \item[\textbf{Q3}] 
    Which fields are most important for the dataset retrieval task?
\end{itemize}

We first obtain features of schema labels as described in Section \ref{slg} and the number of latent factors is set to 40. Then we train a Random Forest with the learned schema label features. The scikit-learn implementation of Random Forest\footnote{\url{http://scikit-learn.org/stable/modules/generated/sklearn.ensemble.RandomForestClassifer.html}} is used with default parameters except the number of trees is set to 25.
In practice, we could choose any multi-label classifier. For each column, we select the top 10 generated schema labels and filter those with probability lower than 0.5. 
For each dataset, we index the generated schema labels as an additional field. 
Table \ref{ndcg} summarizes the NDCG at $k$ and Precision at $k$ of different models.  
Note that, for Schema Label Mixed Ranking (SLMR), we trained three different models and the weights of used fields were forced to be non-zero in order to study the proposed research questions. The weights of used fields for multifield document representation are also set non-zero when optimizing the parameters.

From the results of single-field document ranking, we can see that only utilizing the data table for ranking leads to the worst performance. Scoring on the concatenation of title and description achieved the best results, which indicates that title and description are more important than the data table for ranking a dataset (\textbf{Q3}). Treating all fields of a dataset as a single-field document provides  performance between the previous two models. This result is expected since the length of data tables are usually much larger than titles and descriptions, and therefore dominate the table representation.

By comparing the results of single-field and multifield document ranking, we observe that the combination of the scores of data table, title and description could improve NDCG@k. Though NDCG@k decreases when k increases, the relative improvement against single-field document ranking are more significant. In contrast, for Precision@5, Precision@10, single-field document ranking performs better than multifield document ranking, though the differences are small.  So for \textbf{Q1}, under the setting of multifield document ranking, the content of the data table could help NDCG, but not help Precision of dataset retrieval results.

Without scoring data tables, our proposed schema label mixed ranking approach achieves the highest NDCG on all the rank cut-offs, which indicates that the generated schema labels can be useful to improve the NDCG of dataset retrieval results (\textbf{Q2}).  Though Precision@20 of multifield document ranking are higher than our proposed model, the difference is no more than 0.4\% ($p\_value > 0.9$). 
Significantly, our model outperforms by 21.3\% for Precision@5 ($\frac{0.5-0.4122}{0.4122}$) and by 20.1\% for Precision@10 ($\frac{0.4388-0.3652}{0.3652}$) than the best baseline methods ($p\_value < 0.01$).  Whether data tables are scored or not, Precision@k is not significantly different for schema label mixed ranking. Therefore, under the setting of schema label mixed ranking, data tables make little contribution in this scenario (\textbf{Q1}). One possible reason could be that data tables collected from data.gov contain large quantities of numerical values and will rarely be used to match user queries.

If a schema label mixed ranking model scores only on titles and descriptions ($w_l=0$), it is equivalent to single-field ranking model scoring on titles and descriptions. Therefore, we can compare the results in first and fifth rows in Table \ref{ndcg}. With generated schema labels, the ranking model can have a higher performance on dataset retrieval task (\textbf{Q2}).

\subsection{Schema Label Generation Enhanced Search for Web Tables}

The task of dataset search is similar to Web table search since both tasks use table structure to represent data. The difference is that a large amount of Web tables are entity focused and contain many named entities that can be linked to a knowledge base. However, our datasets collected from the data.gov data portal contain few useful entities in the table. Therefore, a lot of methods designed for Web table ranking cannot be applied to dataset search. The semantic table
retrieval (STR) method proposed by Zhang and Balog \cite{Zhang:2017:ESA} relies on features from knowledge bases (bag of entities)  which are not generally available for the scenario of dataset search. However, the schema label generation based method can be applied to table search. Thus, we performed additional experiments to show the performance of our method for the table search scenario. 

We first generate schema labels for the table corpus shared by Zhang and Balog \cite{Zhang:2017:ESA} using the method proposed in Section \ref{slg}. 
Then we append five additional features to their proposed features\footnote{\url{https://github.com/iai-group/www2018-table/tree/master/feature}} based on schema labels. Each feature is one type of semantic similarity between query and schema labels. Four features are calculated using the measurement proposed by Zhang and Balog (one early fusion feature, three late fusion features) and the last feature is the negative of Word Mover's Distance. 
Finally, like Zhang and Balog, we use Random Forest to perform pointwise regression and the final reported results are averaged over five runs of 5-fold cross-validation and shown in Table \ref{tab:table_rs}. 

\begin{table}[t]
\small
\centering
\caption{Supervised ranking results on table retrieval.}
\label{tab:table_rs}
\begin{tabular}{@{}lllll@{}}
\toprule
Method & NDCG@5 & @10 & @15 & @20 \\ \midrule
STR \cite{Zhang:2017:ESA} & 0.6366 & 0.6571 & 0.663 & 0.6632 \\
Schema Label Features & 0.4489 & 0.5201 & 0.534 & 0.5347 \\
STR + Schema Label Feat. & \bf 0.6530 &\bf 0.6728 &\bf 0.6789 & \bf 0.6761 \\ \bottomrule
\end{tabular}
\end{table}

We can see that schema label features along cannot outperform STR. But combining them results in improvement.
However, by calculating the normalized feature importance measured in terms of Gini score, we find that for STR with schema label features, WMD based measurement contributes the most among all the semantic features. Thus it demonstrates that the schema labels can be valuable for the table retrieval task as well.

Notably, in this table corpus, many tables lack much table content but contain rich text descriptions, which could be unfair for schema label generation-based methods. While for dataset search, each table has values but may lack high quality dataset descriptions. We believe that our schema label generation method can outperform STR in the scenario where text descriptions provide less useful information than the table itself.

\begin{table}[t]
\small
\centering
\caption{Unsupervised ranking results on table retrieval.}
\label{tab:table_rs_unsup}
\begin{tabular}{@{}lllll@{}}
\toprule
Used Fields                                & NDCG@5          & @10             & @15             & @20             \\ \midrule
text                                 & 0.3724          & 0.3891          & 0.4009          & 0.4178          \\
text + data table                    & 0.3901          & 0.4042          & 0.4422          & 0.4686          \\
\begin{tabular}{@{}c@{}}text + data table +   generated labels\end{tabular}
 & \textbf{0.4006} & \textbf{0.4118} & \textbf{0.4495} & \textbf{0.4766} \\
\begin{tabular}{@{}c@{}}text + data table +   original labels\end{tabular}
  & 0.3930          & 0.4055          & 0.4457          & 0.4709          \\
text + original labels               & 0.3785          & 0.3934          & 0.4110           & 0.4283          \\
text + generated labels              & 0.3808          & 0.3955          & 0.4064          & 0.4197          \\ \bottomrule
\end{tabular}
\end{table}

We also show unsupervised ranking results with Equation \ref{mix} in Table \ref{tab:table_rs_unsup}. Unlike Zhang and Balog \cite{Zhang:2017:ESA}, we consider page title, section title and caption as a single text field, in order to reduce the number of hyperparameters (field weights). The results show that generated labels are more effective than original labels for table ranking. It is unsurprising because generated labels often include not only original labels but also additional labels that can benefit the ranking model. We also notice that including the data table field achieves better results than not scoring it, which is contrary to the results of dataset ranking. It is also expected since WikiTables are entity-focused and include a lot of text information while data tables from data.gov include more numeric values.  

\section{Conclusion}

In this paper, we have proposed a schema label enhanced ranking framework for dataset retrieval. The framework has two stages: in the first stage, a schema label generator is trained to generate additional schema labels for each dataset column; in the second stage, given a user query, datasets are ranked by their original fields together with generated schema labels.  
Schema label generation is treated as a multi-label classification task in which each column of a dataset is associated with multiple schema labels. Instead of using hand-curated features, we learn the latent feature representations of schema labels by a CoFactor model in which the dataset-schema label interactions and schema label-schema label interactions are captured.  With the schema label mixed ranking model, the traditional ranking scores for text fields (title, description, data rows) and word embedding-based scores for generated schema labels can be used to rank the datasets. 


We created a new benchmark to evaluate the performance of dataset retrieval. 
The experimental results demonstrate our proposed framework can effectively improve the performance on the dataset retrieval task. It achieved the highest NDCG on all the rank cut-offs compared with all baseline methods.
We also apply our method to the web table retrieval task which is similar to dataset search and find that the features generated from schema labels can help in supervised ranking as well.

\subsection*{Acknowledgment}
This material is based upon work supported by the National
Science Foundation under Grant No.\ IIS-1816325.		
		
\newpage 

\bibliographystyle{splncs04}
\bibliography{acmart}

\end{document}